\documentclass[aps,prl,twocolumn,showpacs,groupedaddress]{revtex4}
\usepackage{amsmath}
\usepackage[latin1]{inputenc}
\usepackage[dvips]{graphicx} 
\usepackage{amssymb}
\usepackage{color}
\usepackage{float}
\newcommand{\T}{{\cal T}}

\newcommand{\V}{{\cal V}}

\begin{document}
\title{Robust quantum coherence above the Fermi sea}

\author{S. Tewari$^{1,2}$, P. Roulleau$^1$, C. Grenier$^3$, F. Portier$^1$, A. Cavanna$^2$, U. Gennser $^2$, D. Mailly $^2$, and P. Roche$^1$}\email{patrice.roche@cea.fr}
\affiliation{$^1$Nanoelectronic group, Service de Physique de l'Etat Condens{\'e}, CEA/IRAMIS/SPEC, CNRS UMR 3680, 91191 Gif-sur-Yvette, France}
\affiliation{$^2$Phynano team, Laboratoire de Photonique et Nanostructures,  CNRS, Route de Nozay, F-91460 Marcoussis, France}
\affiliation{$^3$ Institute for Quantum Electronics, ETH Zurich, 8093, Zurich, Switzerland}
\date{\today}
\begin{abstract}
In this paper we present an experiment where we measured the quantum coherence of a quasiparticle injected at a well-defined energy above the Fermi sea into the edge states of the integer quantum Hall regime. Electrons are introduced in an electronic Mach-Zehnder interferometer after passing through a quantum dot that plays the role of an energy filter. Measurements show that above a threshold injection energy, the visibility of the quantum interferences is almost independent of the energy. This is true even for high energies, up to 130~$\mu$eV, well above the thermal energy of the measured sample. This result is in strong contradiction with our theoretical predictions, which instead predict a continuous decrease of the interference visibility with increasing energy. This experiment raises serious questions concerning the understanding of excitations in the integer quantum Hall regime.
\end{abstract}
\maketitle

A new type of quantum device, relying on the one dimensional edge states of the Quantum Hall regime, where electrons mimic the photon trajectory of a laser beam, has opened a route towards electron quantum optics and manipulation of single electron excitations \cite{Ji03Nature422p415,Feve07Science316p1169,Bocquillon13Science339p1054}. Pauli statistics and interactions provide new ingredients for the physics of the electrons which are not relevant for photons.  For example, when electrons are injected above the Fermi sea, it is fundamental to understand how their phase coherence will be affected by the injection energy. We explore this issue by first using a quantum dot to inject the carriers at a controllable energy into an edge state. Then an electronic Mach-Zehnder interferometer is used to monitor the quantum coherence of the electronic quasiparticle. We find that above a certain threshold the coherence is energy-independent; it is even preserved at energies fifty times larger than the electronic temperature. This is remarkable, since from simple considerations based on Fermi's golden rule, one would expect that the relaxation rate increases with the injection energy, thus reducing quantum coherence. Indeed, our simulations using recent theories \cite{Levkivskyi08PRB78n045322} predict a continuous trend of increasing relaxation. While the origin of this coherence robustness remains unidentified, it has a significant bearing for the implementation of quantum information encoded in electron trajectories \cite{Beenakker03PRL91n147901,Samuelsson04PRL92n026805}.

The edge states used in this new type of devices for electronic quantum optics are obtained by applying a high magnetic field perpendicular to a high mobility two-dimensional electron gas.
When the number of electrons per quantum of flux (the filling factor) is an integer, the transport occurs through one-dimensional channels located at the
edge of sample. The electron motion in these wires is chiral: the electrons drift in one direction with a speed of the order of 10$^4$ to 10$^6$ ms$^{-1}$ \cite{Ashoori92PRB45r3894,Kumada11PRB84n045314}, 
\begin{figure}[H]
\includegraphics[angle=0, trim= 2cm 1cm 2.5cm 0.5cm, width=8.4cm,keepaspectratio,clip]{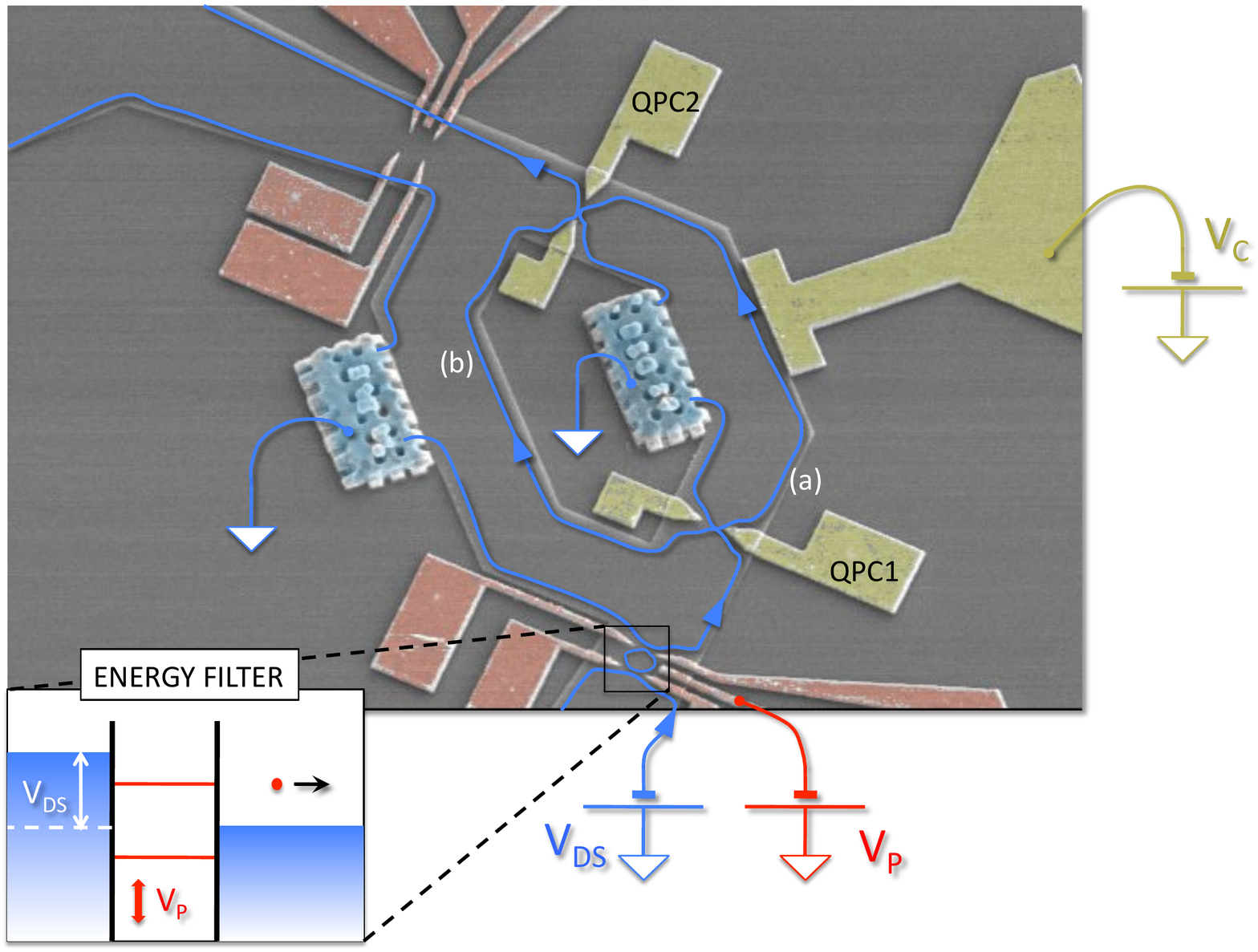}
\caption{\textbf{Measured device. }Colorized scanning electron microscope view of the sample before the final fabrication step in which the gates and ohmic
contacts are connected to larger connecting pads. The sample consists of a quantum dot in series with an electronic Mach-Zehnder Interferometer (MZI).
The Quantum Point Contacts (QPCs) and gates controlling the dot are in red while those controlling the MZI are in yellow. QPC1 and QPC2 serve as the
two beam splitters of the MZI. Note the small ohmic contacts in blue connected to the ground, which prevents spurious quantum interferences inside the interferometer. The chiral trajectory of the excitations in the outer edge state is schematically represented in blue, and the
quantum interference  takes place between the trajectories (a) and (b).   The
mesa is 1.2~$\mu$m wide, each arm of the Mach-Zehnder interferometer is 7.2~$\mu$m long, and the distance between the dot (the energy filter) and the
MZI is 2.8~$\mu$m. $V_P$ is the potential applied on the plunger gate of the dot, allowing a control of the relative position of the energy levels of the dot compared to the Fermi
level. The quantum interferences in the MZI are revealed by sweeping the gate voltage $V_C$, which modifies the trajectory length difference between (a) and (b).}
\label{Sample}
\end{figure}
 thus compensating the confining electric field with the Lorentz force. 
Much recent progress has been made in the understanding of decoherence mechanisms and the energy exchanges
in the integer quantum Hall regime at filling factor two. In this regime, there are two parallel, adjacent channels at each edge in the sample. 
Coulomb interaction has been shown to play a key role in these systems: (i) the quantum coherence at finite temperature is limited by the thermal charge noise of the environment \cite{Roulleau08PRL100n126802,Roulleau08PRL101n186803} (ii) there is an energy relaxation in
out-of-equilibrium edge states and an energy transfer between the two edge states \cite{Lesueur10PRL105n056803} (iii) the visibility of quantum interferences in electronic Mach-Zehnder interferometers exhibits a side lobe structure at finite bias, which
is explained by a beating effect between a neutral and a charged excitation shared by the two edge states \cite{Neder06PRL96p016804,Roulleau07PRB76n161309,Litvin07PRB75n033315,Levkivskyi08PRB78n045322,Huynh12PRL108n256802}. Here, we inject a quasi particle at a well-defined energy above the Fermi sea into an edge channel.  We then explore to which extent such a single charge behaves as a free non-interacting particle in this interacting quantum system, thus directly probing the validity of the Landau's Fermi liquid picture \cite{Landau57JETP3p920,Landau59JETP8p70}. While extensively studied in diffusive
quantum conductors, this question has never been addressed experimentally for the case of one dimensional chiral conductors.  We note, however, that this is a key point for  quantum information experiments using electrons transported through edge channels. Experimentally, we use a quantum dot (QD) as an energy filter, and a Mach-Zehnder interferometer to probe the injected quasi-particle's loss of phase coherence. 
 \begin{figure}
\includegraphics[angle=0,trim= 4.5cm 1cm 5cm 1cm,width=8.5cm,keepaspectratio,clip]{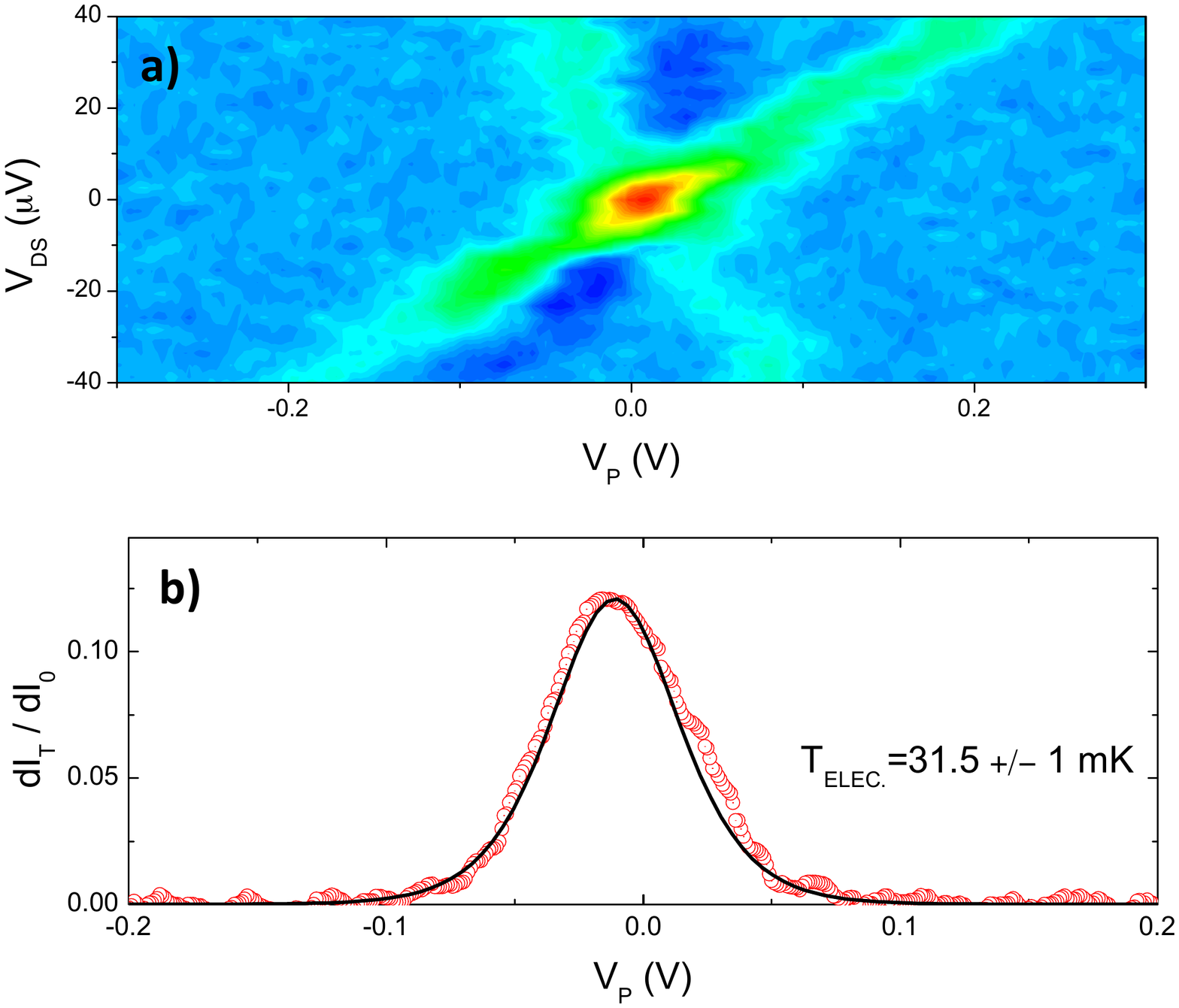}
\caption{a) Color plot of the transmission probability through the quantum dot as a function of the plunger voltage V$_P$ and
the drain source voltage V$_{DS}$. \textbf{b)} Transmission probability as a function of V$_P$ at zero bias. The continuous line is a fit 
of this resonance.}
\label{Qdot}
\end{figure}

\begin{figure}
\includegraphics[angle=0, trim= 1.5cm 0.5cm 2cm 0cm, width=8.5cm,keepaspectratio,clip]{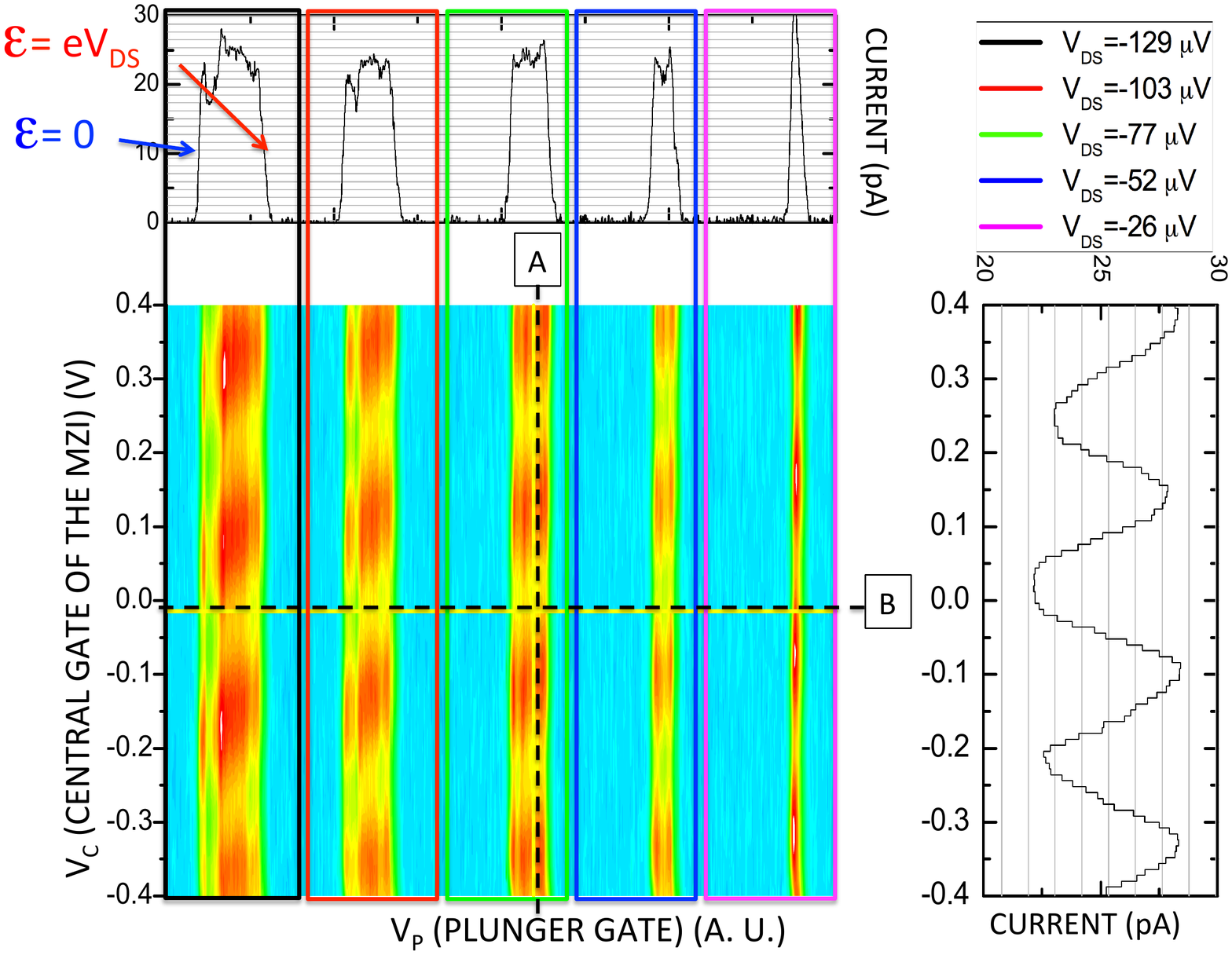}
\caption{Current as a function of the plunger gate voltage $V_P$ and the central gate voltage $V_C$ of the MZI  for
different values of the drain source bias $V_{DS}$. The top graph, representing the current as a function of $V_P$, corresponding to a 2D plot cut through line B. The right graph shows the interference pattern as a function
of $V_C$ corresponding to a cut trough line A.}
\label{Result1}
\end{figure}

The QD consists of two quantum point contacts (QPC) and a plunger gate (see Figure \ref{Sample}). Figure \ref{Qdot}a shows  a 2D plot of the transmission probability through the dot as a function of the drain-source bias and the plunger gate voltage V$_P$. The transmission probability is defined as $dI_T/dI_0$, where $I_T$ is the transmitted current through the QD and $I_0$ is the impinging current. From this measurement we deduce the lever arm of the plunger gate $\alpha=d\epsilon/d(eV_P)= 1.46~10^{-4}$, where $\epsilon$ is the QD energy. As an example, figure \ref{Result1} (top graph) displays the current I$_T$ as a function of V$_P$. It is clear that I$_T$ is almost constant for energies varying between 0 and e$V_{DS}$. This signals an absence of excited states for this particular tuning of the quantum dot. We deduced an electronic temperature of 31$\pm$1~mK by fitting the resonance of the transmission probability (see Figure \ref{Qdot}b)
as a function of $V_P$ with $dI_T/dI_0\propto \cosh^{-2}(\delta /2k_BT)$ \cite{Beenakker91PRB44p1646}, where $\delta=\alpha(eV_P-eV_{P_0})$ is the energy difference between the QD energy level and the Fermi level, and $V_{P_0}$ is the plunger gate voltage which maximizes the conductance.
In practice, the presence of excited states in the dot limited the energy range we explored to 120~$\mu$eV, close to the addition energy estimated to be of the order of 150~$\mu$eV (see SM).

 Once the QD is characterized, 
 we measure the current through the whole system, constituted by the QD in series with the MZI, and operated in its optimal parameter regime for maximum visibility (see SM). 
 Probing the decoherence at finite energy is simply realized by measuring 
 the relative amplitude of the current oscillations revealed by sweeping the central gate voltage $V_C$ of the Mach-Zehnder interferometer. A color plot of the current as a function of $V_C$ and $V_{P}$ is displayed in figure
 \ref{Result1}. This plot corresponds to experiments with
 five different drain-source voltages $V_{DS}$.  One would expect a current injection only for a range of $V_P$ values between 0 and $V_{DS}$ (see inset of Figure 1), and this is indeed what can be seen in the top graph of Figure \ref{Result1}, where the window of non-zero current (around 25~pA) increases linearly with increasing source-drain voltage. Sweeping
 $V_C$ while Vp is within this window reveals the quantum interferences. The interference pattern is clearly  observable in Figure \ref{Result1}.
 It is easily noticed that, within the window for non-zero current, the amplitude of the current oscillations appears to change very little; a clear indication that the quantum interferences, and hence the coherence, are more or less independent of the injection energy of the electrons.
 
\begin{figure}
\includegraphics[angle=0, trim= 0cm 0cm 0cm 0cm, width=8.5cm,keepaspectratio,clip]{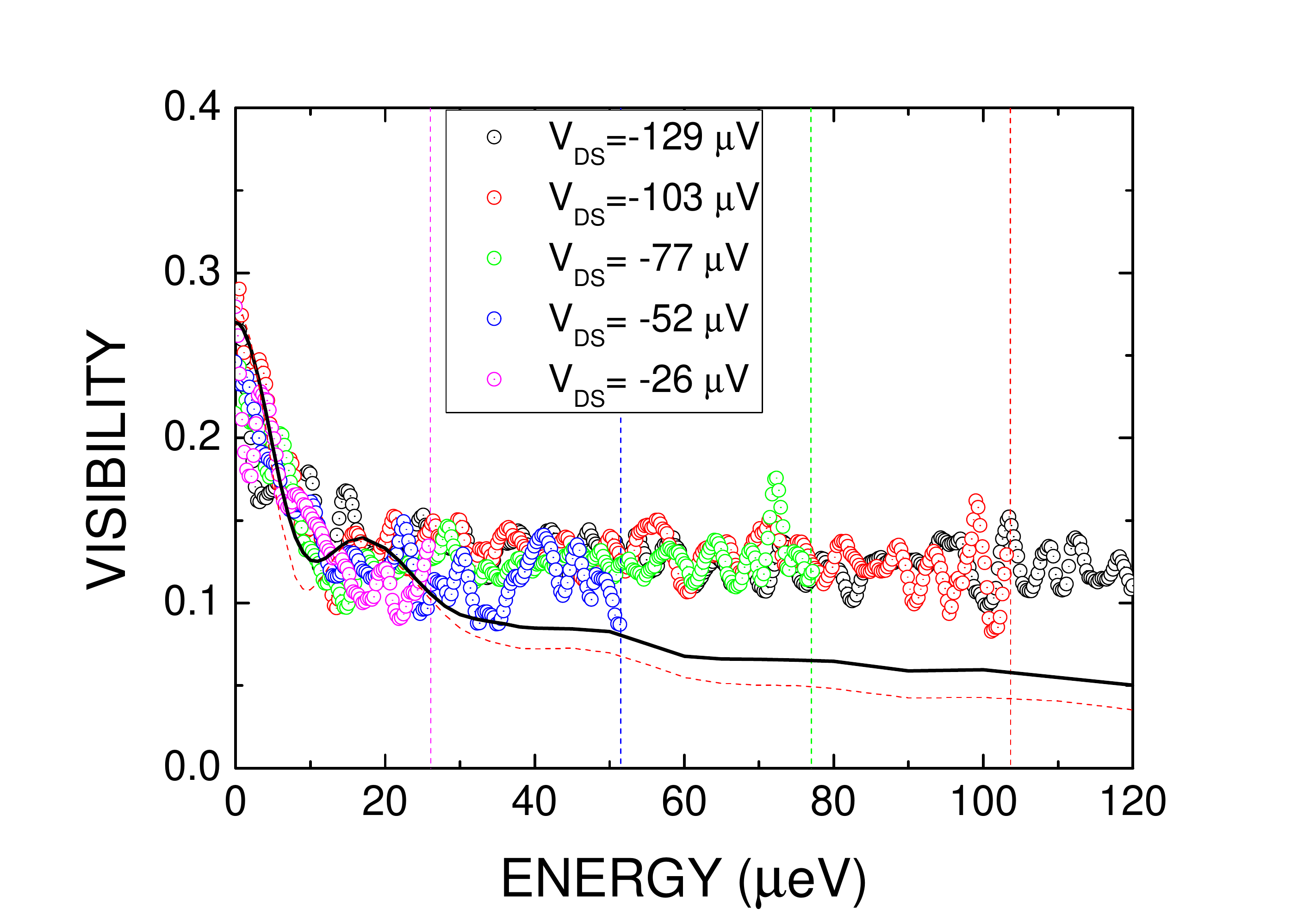}
\caption{Visibility of quantum interferences as a function of the energy for five different bias voltages. The vertical lines 
represent the maximum energy achievable for each bias. The solid and the dashed lines show the calculated visibility for interaction parameters $\theta=\pi/2$ and
$\theta=\pi/3$, respectively. }
\label{Visi}
\end{figure}

To be more quantitative, we plot in Figure \ref{Visi} the visibility $\V=\frac{I_{MAX}-I_{MIN}}{I_{MAX}+I_{MIN}}$ of quantum interferences as a
function of the injected energy. Surprisingly, for energies greater than 20 $\mu$eV, the visibility remains almost constant instead of decreasing down to zero as expected.
We need to consider whether the saturation of the visibility is due to an unexpected robustness of the quantum coherence, or e.g, to the finite distance, 2.7 $\mu$m, between the energy filter and the MZI.
Indeed, a short relaxation length ($\ll$~ 2.7$\mu$m) would mean that the electrons injected above the Fermi sea are fully relaxed before reaching the entrance of the interferometer. However, this scenario is
in contradiction both with recent experimental results and theories, as we will now show. 

On the experimental side, experiments at filling factor two with a hot electron distribution in one channel have demonstrated that the inelastic length is of the
order of 3~$\mu$m \cite{Lesueur10PRL105n056803}. This length is in fact longer than the separation between the QD
and the MZI. 
A second experimental evidence against fast relaxation is the Hong Ou Mandel (HOM) experiment with electrons injected one by one into edge states via quantum dots, which are excited with a periodic radio frequency gate voltage matched to their addition energy \cite{Bocquillon13Science339p1054}. The observed HOM dip, resulting from the collision of two electrons emerging from two separate sources, implies a coherent propagation of the injected electrons. This observation, for which electrons have been injected at energies higher than 1~meV, shows qualitatively that an important fraction of the injected electrons remains
coherent, as observed in our experiment.

Furthermore, our quantitative model calculations show that the electrons reaching the MZI are only partially relaxed. 
We describe our system through an approach similar to \cite{Degiovanni09PRB80n241307}, which considers the relaxation of an energy resolved single electron wavepacket  $\phi_{k_0}(x)\propto e^{ik_0x}$ injected in the outer edge channel, and takes into account the coupling of the $\nu=2$ edge channels \cite{Levkivskyi08PRB78n045322}. This approach has proved useful for the understanding of unexpected phenomena observed on out-of-equilibrium MZIs \cite{Huynh12PRL108n256802}, and of the observed energy transfer between edge channels\cite{Degiovanni10PRB81n121302,Lesueur10PRL105n056803}. More specifically we calculated,  as a function 
of the normalized propagation length $k_0L$, the elastic transport probability ${\cal Z}(k_0L)$ and the diagonal elements of the density matrix $\delta n(k,k_0L)$ that account for the relaxation of the injected electron
on lower modes. To include the relaxation on the length $L_{QD}$ separating  the QD from the MZI
we computed $\delta n(k,k_0L_{QD})$ and $ \V_{Theo.}\propto \int \delta n(k,k_0L_{QD}){\cal Z}(2k_0L_{MZI}) dk$, where $L_{MZI}$ is the MZI arm length. The results
are plotted in figure \ref{Visi} for two different values of the coupling parameter $\theta$ measured in recent experiments \cite{Inoue14PRL112n166801,Bocquillon13NatCom4n1839}. 
In order to fit our data and  the
decrease of the visibility at low energy we chose a drift velocity of 5$\times$10$^{4}$~ms$^{-1}$.
 As expected, but in contradiction with our observations, the theory
does not lead to a saturation of the visibility at high energy, but to a decrease with a slope less pronounced than at low energy. This theoretical
result reflects the fact that wave packet is partially (and not totally) relaxed before entering the MZI. 

The robustness of the quantum coherence, which we have demonstrated here, increases the attraction of electron guns for quantum optics type experiments.  At the same time, our results raise important questions regarding our understanding of excitations in the integer quantum Hall regime.  To further elucidate the relaxation in edge states, the next step to this experiment will be to implement a second quantum dot after the MZI, to make the spectroscopy of the injected electron after its propagation through the interferometer.

\textit{acknowledgment} PR would like to thank Eugene Sukhorukov for stimulating discussions. This work
has been supported by the French ANR contract  11-BS04-022-01 IQHAR.

\pagebreak

\begin{center}
\textbf{\large Supplemental Materials: Robust quantum coherence above the Fermi sea}
\end{center}
\setcounter{equation}{0}
\setcounter{figure}{0}
\setcounter{table}{0}
\setcounter{page}{1}
\makeatletter
\renewcommand{\theequation}{S\arabic{equation}}
\renewcommand{\thefigure}{S\arabic{figure}}
\renewcommand{\bibnumfmt}[1]{[S#1]}
\renewcommand{\citenumfont}[1]{S#1}

\section{Experiment}
This section provides a short description of the experimental procedure.
The tuning of the quantum dot (QD) and the Mach-Zehnder interferometer (MZI) is realized in several steps. We first characterize the
four quantum point contacts (QPCs) constituting the barriers of the QD and the beam splitters of the Mach-Zehnder interferometer. We also check
the quality of the small ohmic contact inside the MZI. 

\begin{figure}[h]
\includegraphics[angle=0,width=8.5cm,keepaspectratio,clip]{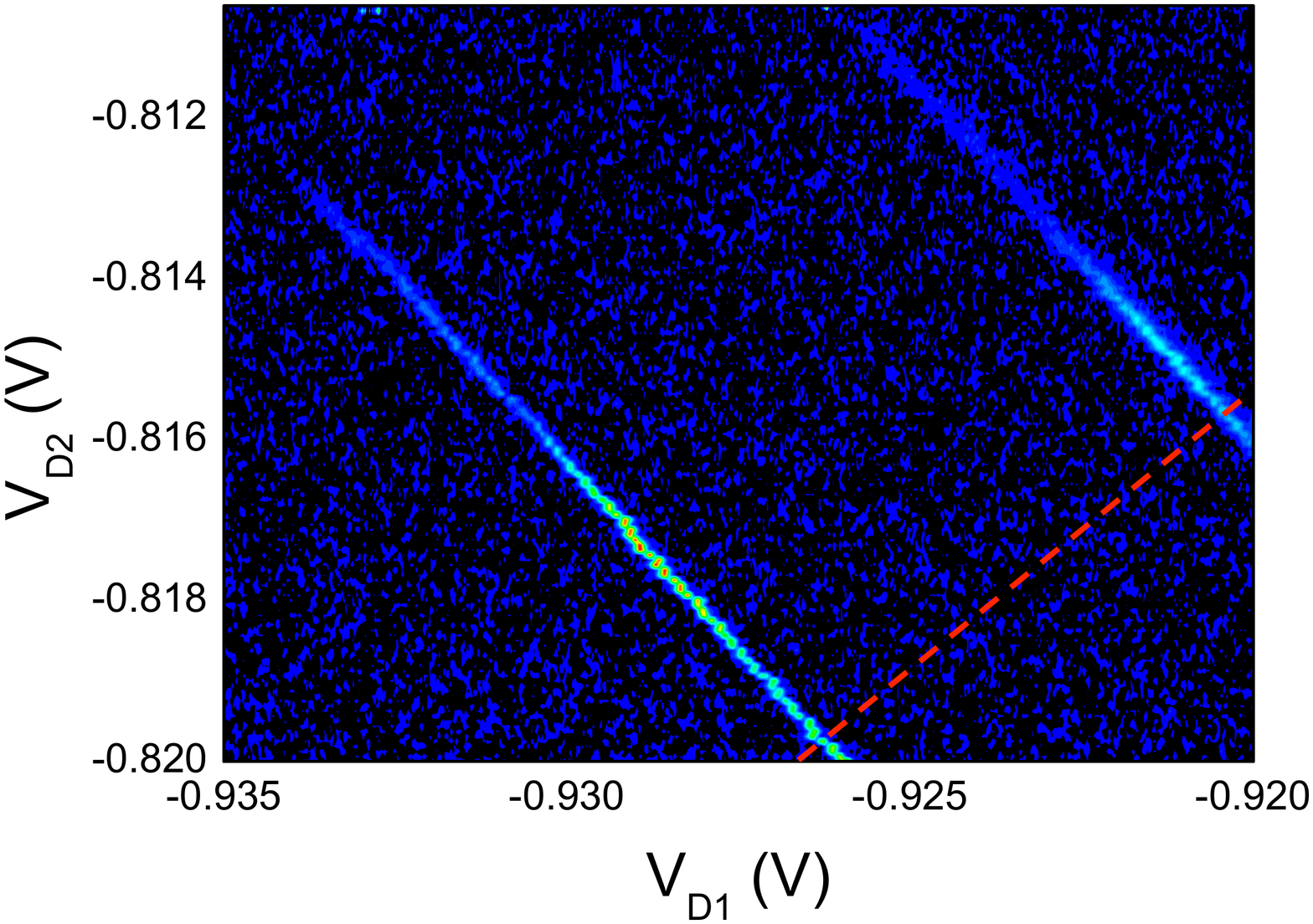}
 \caption{2D color plot of the transmission probability dI$_T$/dI$_0$ through the quantum dot as a function of the bias voltage applied on the two QPCs D1 and D2 defining the quantum dot. The resonant conditions in the conductance are characterized by colored segments in the 2D plot. The conductance measurement following the red dash line is plotted in figure \ref{fig:SM2}.}
 \label{fig:SM1}
\end{figure}

To tune the quantum dot (QD) we first realize  a 2D plot of the conductance through the QD as a function of V$_{D1}$ and V$_{D2}$,
the gate voltages applied to the two QPCs D1 and D2 defining the QD. Near pinch-off, one can observe  non-zero conductance segments in the 2D plot resulting from the alignement of the QD energy levels with the Fermi energy. 
The conductance through the quantum dots is usually stable for a day, after which sudden events - probably due to charges trapped in the vicinity of the QD - modify its  tuning. This limits our possible data acquisition time.
 Note that occasionally we  observe
excited states in an energy range of the order of 40~$\mu$eV. As we want to inject a particle at a defined energy, we always
carefully tune the QD such that there are no such excited states in the energy range explored. Although the visibility for V$_P$
sweeps seems to be reduced when the current
through the quantum dot increases because of the presence of an excited state,
 a general study of this behavior is beyond  the scope of the present paper.

\begin{figure}
\includegraphics[angle=0,width=8.5cm,keepaspectratio,clip]{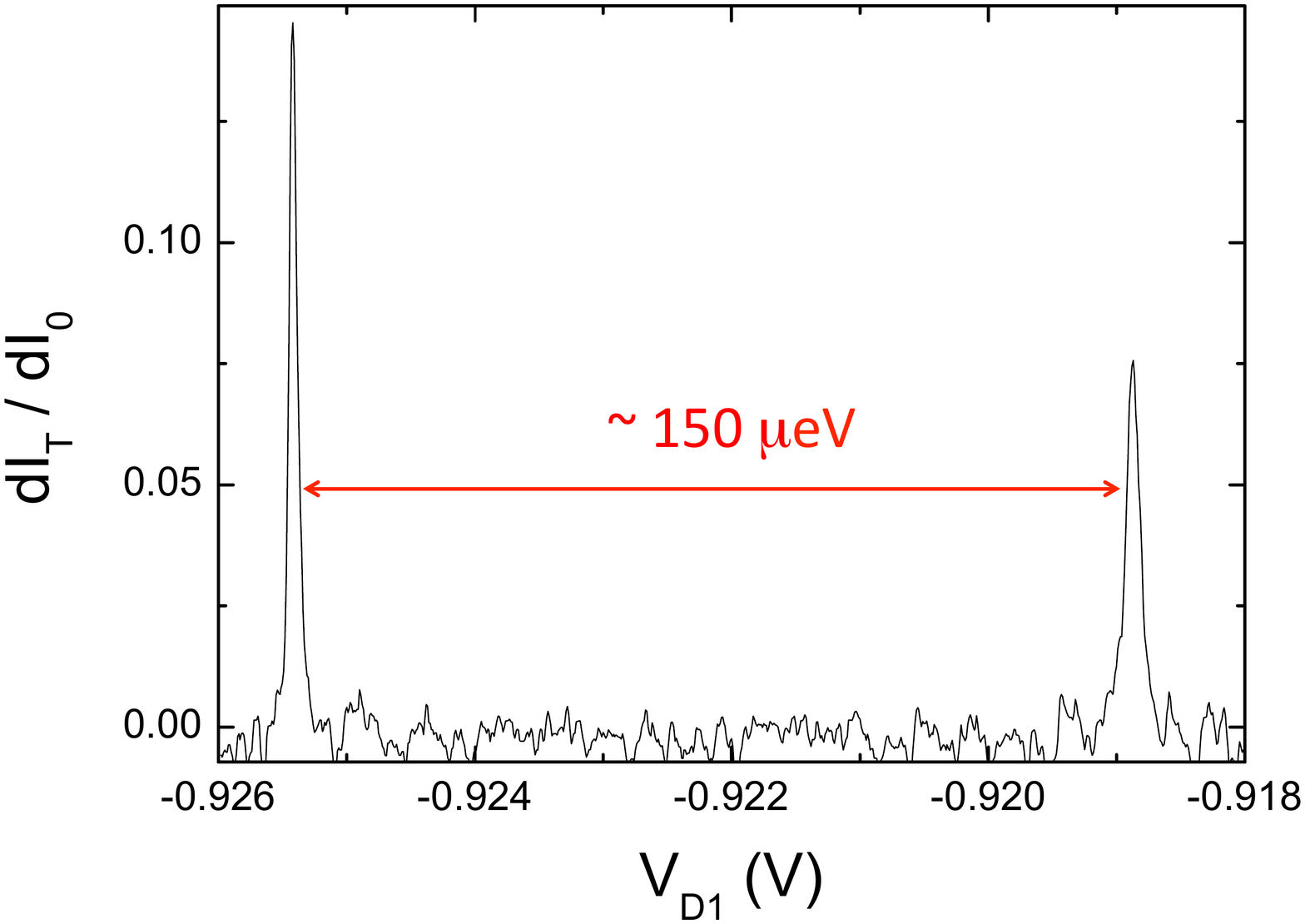}
 \caption{Measured conductance when varying V$_{D1}$ and V$_{D2}$ following the red dash line of figure \ref{fig:SM1}. The ratio between the
 width of the peaks and distance between the two peaks is the ratio between the temperature and the addition energy. Knowing the actual electronic temperature 
 (see main text) we infer an addition energy of the order of 150~$\mu$eV. }
 \label{fig:SM2}
\end{figure}

The addition energy of the order of 150~$\mu$eV is estimated by plotting the conductance following
the red dotted line on the 2D plot of figure \ref{fig:SM1}. The measured conductance is displayed in figure \ref{fig:SM2}.  The ratio between the width of
the peaks and the peak separation is equal to the ratio between the temperature and the addition energy. Note that we did not succeed
in measuring the addition spectrum by varying the plugger gate bias, most probably because the cross talk between the plunger gate
and the QPCs defining the barriers of the quantum dot leads to a detuning when varying V$_P$ on a large range. 

We also checked that the capacive coupling between QPCs defining the MZI and the the Quantum Dot do not alter its tuning. The maximum visibility of quantum interferences through the MZI is observed when the transmissions $\T_1$ and $\T_2$ of the MZI beam splitters QPC1 and QPC2 are equal to 1/2 \cite{Roulleau07PRB76n161309}. We roughly set the  QPC1 and QPC2 to half transmission using the former characterization of the QPCs, and
we then fine-tune the QPCs in order to obtain the highest visibility of quantum interferences when sweeping V$_C$.

\section{Theory}

This section provides a short description of the 
 theoretical approach on which rely our test with the experimental observations
 presented in the main text. Notably, it contains the necessary ingredients to 
 produce the curves compared to the data in Fig. 3 of the main text.
 
\subsection{Hamiltonian and bosonization of the two channel system}

\subsubsection{Statement of the problem}

The computation of the elastic transport probability
relies on a description of the edge channel as a one-dimensional
chiral electron fluid with linear dispersion : $\varepsilon (k) = \hbar v_F k$. In this picture, 
assuming a short range interaction, the hamiltonian describing the evolution of the electrons travelling within the system made out of the two edge channels reads~:

\begin{eqnarray}
 \label{eq:hamiltonian_fermions}
 \mathcal{H} =& \int dx \sum_{\alpha=1,2} \psi_\alpha^\dagger(-iv_F\partial_x)\psi_\alpha(x) \nonumber\\
 {} +&\int dx \sum_{\alpha,\beta} \psi_\alpha^\dagger(x)\psi_\alpha(x)U_{\alpha\beta}\psi_\beta^\dagger(x)\psi_\beta(x)\,.
\end{eqnarray}

\subsubsection{Equations of motion and plasmon scattering}

In the integer quantum Hall regime, the two edge channels are described by two chiral bosonic fields $\phi_{\alpha}(x,t)$ related to the charge densities
via $\rho_\alpha (x,t) = -\frac{e}{\sqrt{\pi}}\partial_x\phi_\alpha(x,t)$~\cite{Wen:1990-1}.
Our work closely follows the Luttinger liquid approach presented in \cite{degiovanni-2009,Levkivskyi:2008-1,Chalker:2007-1,PhysRevLett.111.136807} . The equations of motion for the two boson fields are:
\begin{eqnarray}
 \left( \partial_t + v_1 \partial_x \right)\phi_1(x,t) & = & U\phi_2(x,t)\\
 \left( \partial_t + v_2 \partial_x \right)\phi_2(x,t) & = & U\phi_1(x,t)\,,
\end{eqnarray}
where the two velocities $v_{1,2}$ take into account the intrachannel interactions.\\
Assuming that the coupling between the two edge channels acts over a finite length $L$, going to Fourier
space $\tilde{\Phi}_\alpha(x,\omega) = \int dt\, e^{i\omega t}\Phi_\alpha(x,t)$ provides the solution for the outgoing fields $\tilde{\Phi}_\alpha(L,\omega)$ as
a linear combination of the input ones $\tilde{\Phi}_\alpha(0,\omega)$. The coefficients $\mathcal{S}_{\alpha\beta}(L,\omega)$ of this linear expansion
constitute the coefficients of the plasmon scattering matrix, which encode the effect of the propagation along a finite length $L$ under the action of interactions.\\
In the case of the short-range coupling between two edge states, this matrix can be written in the general form~:
\begin{equation}
\label{eq:plasmon_scatt}
 \mathcal{S}(L,\omega) = e^{i\omega L/\bar{v}}\cdot \exp{\left[-i\omega L/v(\cos{\vartheta}\sigma^z+\sin{\vartheta}\sigma^x)\right]}\,.
\end{equation}
In~\eqref{eq:plasmon_scatt}, $\bar{v}$ and $v$ are such that the velocities of the eigenmodes are given by $v_\pm^{-1} = \bar{v}^{-1} \pm v^{-1}$.
Of more interest is the angle $\vartheta$ defined by~:
\begin{equation}
 \sin{\vartheta} = \frac{U}{\sqrt{U^2+\left(\frac{v_1-v_2}{2}\right)^2}}\,,
\end{equation}
which represents the strength of interactions. The limits $\vartheta \rightarrow 0$ correspond to a vanishing coupling, whereas $\vartheta \rightarrow \pi/2$ indicates a strong coupling. Experimentally,
values of this mixing angle between $\pi/3$~\cite{PhysRevLett.112.166801} and $\pi/2$~\cite{altimiras2009non,bocquillon2013separation} have been reported, indicating a strong coupling between the edge channels.\\

This plasmon scattering picture of interaction in integer quantum Hall edge channels has been used in~\cite{degiovanni-2009-24} to describe the generic coupling of an edge channel to 
a linear electromagnetic environment, has provided a satisfactory explanation to the energy exchange between edge channels~\cite{degiovanni-2009,altimiras2009non} and qualitatively reproduces the dependence of the visibility of the  interference of MZI with bias voltage \cite{Levkivskyi08PRB78n045322,Huynh12PRL108n256802}. Finally, the plasmon scattering matrix
can be shown to be related to the finite frequency admittance matrix. The measurement of the latter in the case of interacting edge channels in the $\nu = 2 $ regime has been presented in~\cite{bocquillon2013separation}.\\
In our work, we make use of this approach to compute the visibility of the interference fringes of an electronic Mach-Zehnder interferometer.

\subsection{MZI visibility}
\begin{figure}
\includegraphics[angle=0,width=8.5cm,keepaspectratio,clip]{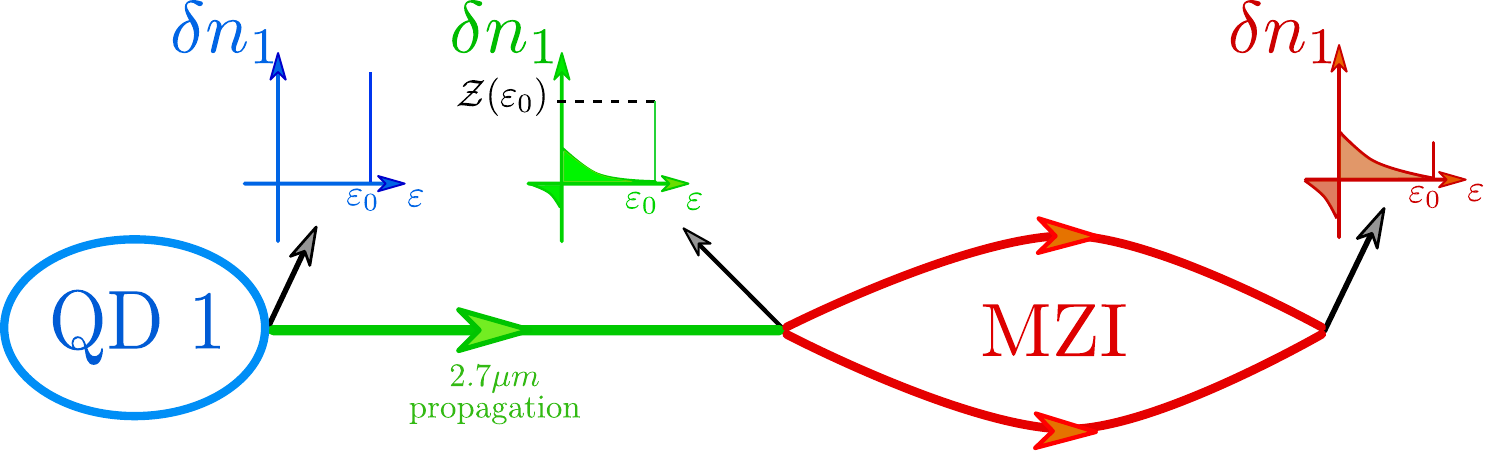}
 \caption{Sketch of the experimental setup. The initial distribution (in blue) originating from the quantum dot QD1 first propagates in the $2.8 \mu$m region (green), generating
 the out of equilibrium distribution $\delta n_1$ (in green), which is the input distribution of the Mach-Zehnder interferometer (red). The visibility is finally measured on the output of the latter.
}
 \label{fig:RSketch}
\end{figure}

Fig.~\ref{fig:RSketch} depicts the system under consideration : a quantum dot emitting an energy resolved electron is immediately followed by two coupled edge channels,
which precede the Mach-Zehner interferometer whose visibility is of interest to us.

Our evaluation of the visibility takes into account the propagation through the region just before the MZI, and the interactions within the MZI itself. 
Thus, the first step is to evaluate the distribution after the region immediately after the dot, assuming an initial distribution of the form (energy resolved emission, blue sketch in Fig.~\ref{fig:RSketch})~:
\begin{equation}
 \delta n(k) = \delta (k-k_0)\,.
\end{equation}
Then, the nonequilibrium distribution $\delta n_1(k)$ after the propagation in the region immediately after the dot and before the first QPC of the MZI (green sketch in Fig.~\ref{fig:RSketch})
is obtained with the plasmon scattering approach sketched in the previous section. This distribution is the entrance electronic population of the MZI and contains exactly one extra electron : $\int dk \delta n_1(k) = 1$.\\

The second ingredient is to find how the visibility of the interference fringes is affected by the interactions within the MZI. Neglecting the effect of Coulomb interactions
at the entrance and exit QPC's, the visibility of the interference fringes for an energy resolved single electron excitation is shown to be~\cite{Levkivskyi:2008-1}~:
\begin{equation}
 \mathcal{V} \propto \mathcal{Z}(\varepsilon_0=\hbar v_Fk_0)\,,
\end{equation}
where $\mathcal{Z}(\varepsilon)$ is the elastic transport probability (weight of the peak of the red distribution sketched in Fig.~\ref{fig:RSketch}). 
Notably, in the absence of coupling, $\mathcal{Z} \rightarrow 1$ and so does the visibility.\\

Finally, to integrate the effect of the region before the interferometer, one assumes that the electrons entering the MZI relax independently. This is legitimate as long as the MZI never contains more than one additional electron.  This is the main assumption of our model and it is compatible with the
experiment: the dc current is of the order of 25~pA leading to a mean electron spacing of 5~ns for a time of flight through the MZI of the order of 0.1ns. Since the supplementary excitation contains only one electron, neglecting the blocking of the 
phase space available for relaxation seems to be reasonable. Then, within this picture, the visibility becomes~:
\begin{equation}
 \mathcal{V} \propto \int dk \delta n_1 (k) Z(k)\,.
\end{equation}
This formula has been used to produce the curves compared to the experimental data in Fig. 3 of the main text.


\end{document}